\documentclass[preprint,showpacs,preprintnumbers,amsmath,amssymb]{revtex4}
\usepackage{graphicx}
\usepackage{dcolumn}
\usepackage{bm}
\usepackage{amsmath}

\begin{document}


\title{\bf Nuclear Schiff moment in nuclei with
soft octupole and quadrupole vibrations}
\author{N. Auerbach$^{1}$, V.F. Dmitriev $^{2}$, V.V. Flambaum$^{3}$,
A. Lisetskiy$^{4}$, R.A. Sen'kov$^{2}$, and V.G. Zelevinsky$^{5}$}
\address{$^1$ School of Physics and Astronomy, Tel Aviv University, Tel
Aviv, 69978, Israel \\
$^2$ Budker Institute of Nuclear Physics, Novosibirsk 630090, Russia \\
$^{3}$School of Physics, University of New South Wales, Sydney
2052, Australia \\
$^4$ GSI, Theory Department, 64291 Darmstadt, Germany \\
$^{5}$National Superconducting Cyclotron Laboratory and\\
Department of Physics and Astronomy, Michigan State University,
East Lansing, MI 48824-1321, USA\\
\\
}
\date{\today}

\begin{abstract}
Nuclear forces violating parity and time reversal invariance
(${\cal P},{\cal T}$-odd) produce ${\cal P},{\cal T}$-odd nuclear
moments, for example, the nuclear Schiff moment. In turn, this
moment can induce the electric dipole moment in the atom. The
nuclear Schiff moment is predicted to be enhanced in nuclei with
static quadrupole and octupole deformation. The analogous
suggestion of the enhanced contribution to the Schiff moment from
the soft collective quadrupole and octupole vibrations in
spherical nuclei is tested in this article in the framework of the
quasiparticle random phase approximation with separable quadrupole
and octupole forces applied to the odd $^{217-221}$Ra and
$^{217-221}$Rn isotopes. We confirm the existence of the
enhancement effect due to the soft modes. However, in the standard
approximation the enhancement is strongly reduced by a small
weight of the corresponding ``particle + phonon" component in a
complicated wave function of a soft nucleus. The perspectives of a
better description of the structure of heavy soft nuclei are
discussed.
\end{abstract}


\maketitle

\section{Introduction}

The search for interactions violating time reversal (${\cal T}$-)
invariance is an important part of studies of fundamental
symmetries in nature. The manifestations of ${\cal CP}$-violation
(and therefore, through the ${\cal CPT}$-theorem, of ${\cal
T}$-invariance) in systems of neutral $K$- and $B$-mesons
\cite{fleischer02} set limits on physical effects beyond the
standard model. The main hopes for the extraction of
nucleon-nucleon and quark-quark interactions violating fundamental
symmetries emerge from the experiments with atoms and atomic
nuclei, see the recent review \cite{ginges04} and references
therein. For example, the best limits on ${\cal P},{\cal T}$-odd
forces have been obtained from the measurements of the atomic
electric dipole moment (EDM) in the $^{199}$Hg \cite{romalis01}
and $^{129}$Xe \cite{jacobs95} nuclei. As we know from past
experience with ${\cal P}$-odd forces, see the review article
\cite{flamgrib95}, there are powerful many-body mechanisms in
heavy atoms and nuclei which allow one to expect a significant
amplification of effects generated on the level of elementary
interactions. There are also suggestions for using possible
molecular and solid state enhancement mechanisms
\cite{hudson02,lamoreaux02,mukhamedjanov05}.

Theoretical calculations of atomic EDM proceed through the {\sl
nuclear Schiff moment} ${\bf S}$ since the nuclear EDM is shielded
by atomic electrons \cite{FKS84}. The Schiff moment produces the
${\cal P,T}$-odd electrostatic potential that, in turn, induces
the atomic EDM. The expectation value of the vector operator ${\bf
S}$ in a stationary nuclear state characterized by certain quantum
numbers of angular momentum, $JM$, is possible only for $J\neq 0$
owing to the requirements of rotational invariance. Since all
even-even nuclei have zero ground state spin, we need to consider
an odd-$A$ nucleus. Furthermore, the non-zero expectation value of
a polar vector ${\bf S}$ requires parity non-conservation in a
nucleus; in addition, being proportional to the ${\cal T}$-odd
pseudoscalar $\langle ({\bf S}\cdot{\bf J})\rangle$, this
expectation value reveals the violation of ${\cal T}$-invariance.

A reliable evaluation of the nuclear Schiff moment should include
the estimates of renormalization effects due to ``normal" strong
interactions inside the nucleus. The core polarization by the odd
nucleon is important, especially in the case of the odd neutron,
as $^{199}$Hg and $^{129}$Xe. Calculations
\cite{FV,DS03,DS04,dejesus05} show that the resulting
configuration mixing, depending on details of the method, may
change the result of the independent particle model by a factor of
about 2. A possibility of using accidental proximity of nuclear
levels with the same spin and opposite parity was pointed out in
Refs. \cite{FKS84,haxton83}. In such approaches, possible {\sl
coherent} enhancement mechanisms are usually not considered.

The statistical many-body enhancement of parity non-conservation
in the region of the high level density of neutron resonances was
predicted theoretically (see e.g. reviews \cite{SF82,flamgrib95}
and references therein). The existence of such enhancement is now
well documented experimentally \cite{mitchell}. The simultaneous
violation of parity and time reversal invariance can be enhanced
by the firmly established ${\cal P}$-violation due to ${\cal
P}$-odd ${\cal T}$-even weak interactions. The idea of a possible
role of {\sl static octupole deformation} \cite{SF,SF82,FZ95,SA}
exploited the parity doublets which appear in the presence of
pear-shaped intrinsic deformation of the mean field. The doublet
partners have similar structure and relatively close energies so
that they can be more effectively mixed by ${\cal P}$-odd forces.
The Schiff moment in the body-fixed frame is enhanced being in
fact proportional to the collective octupole moment. The
microscopic calculations \cite{AFS96,SAF97} predict a resulting
Schiff moment by two-three orders of magnitude greater than in
spherical nuclei; this enhancement was confirmed  in Refs.
\cite{engel03,DE}. The uncertainties related to the specific
assumptions on ${\cal P,T}$-odd forces and different
approximations for nuclear structure are on the order of a factor
2 for the resulting Schiff moment.

It was suggested in Ref. \cite{hayes00} that {\sl soft octupole
vibrations} observed in some regions of the nuclear chart more
frequently than static octupole deformation may produce a similar
enhancement of the Schiff moment. This would make heavy atoms
containing nuclei with large collective Schiff moments attractive
for future experiments in search for ${\cal P,T}$-violation;
experiments of this type are currently under progress or in
preparation in several laboratories. Recently we performed
\cite{FZ03} the estimate of the Schiff moment generated in nuclei
with the quadrupole deformation and soft octupole mode and showed
that the result is nearly the same as in the case of the static
octupole deformation.

A related idea is explored in the present paper. It is known that
some nuclei are soft with respect to {\sl both quadrupole and
octupole} modes, see for example recent predictions for
radioactive nuclei along the $N=Z$ line \cite{kaneko02}. The light
isotopes of Rn and Ra are spherical but with a soft quadrupole
mode and therefore large amplitude of quadrupole vibrations. The
spectra of these nuclei display long quasivibrational bands
\cite{jarmstrong99} based on the ground state and on the octupole
phonon, with positive and negative parity, respectively. These
bands are connected via low-energy electric dipole transitions.
This situation seems {\sl a-priori} to be favorable for the
enhancement of ${\cal P},{\cal T}$-odd effects.

Below we examine the question whether the  enhancement indeed
exists in some spherical nuclei which have both collective
quadrupole and octupole modes. The main weak interaction mixing
can be expected to occur between the levels of the same spin and
opposite parity that carry a significant admixture of the (${\rm
particle}\; + 2^{+}\; {\rm phonon}$) and (${\rm particle}\; +
3^{-}\; {\rm phonon}$) states with the odd nucleon in the same
single-particle orbit. In the odd-neutron nuclei the Schiff moment
is originated by the proton contribution to the collective phonon.
A number of such nuclei have an appropriate opposite parity level
with the same angular momentum close to the ground state. The
mixing between the states of above mentioned nature can imitate
the evolution of the spherical nucleus to the deformed pear-shaped
state. In analogy with Refs. \cite{AFS96,SAF97}, this might lead
to enhancement of Schiff moments.

We will show that the effect does exist but, in the framework of
the standard quasiparticle random phase approximation (QRPA), it
is essentially compensated by the strong spreading of the
single-particle strength that significantly reduces the weight of
the desired configuration in the odd-nucleus ground state. We
concentrate here only on physics connected to the soft vibrational
modes. At the next stage the task of theory should be to go beyond
the RPA and to combine all effects generated by ${\cal
P,T}$-violating interactions, including the influence of the core
polarization and weak interaction admixtures to the phonon
structure which are not discussed on the present work. Such
effects should be included for the quantitative answer; however
they are not expected to lead to a qualitative enhancement.

\section{Collective Schiff moment in spherical nuclei}

Consider an odd-$A$ nucleus with two close levels of the same spin
$J$ and opposite parity, ground state $|{\rm g.s.}\rangle$ and
excited state $|x\rangle$. The energies of these states are
$E_{{\rm g.s.}}$ and $E_{x}$, respectively. Let $W$ be a (${\cal
P},{\cal T}$)-odd interaction mixing nuclear states. Assuming that
the mixing matrix elements of ${\bf S}$ and $W$ are real, we can
write down the Schiff moment emerging in the actual mixed ground
state as
\begin{equation}
{\bf S}= 2\,\frac{\langle {\rm g.s.}|W|x\rangle \langle x|{\bf
S}|{\rm g.s.}\rangle}{E_{{\rm g.s.}}-E_{x}}.   \label{2.1}
\end{equation}
However, as it was explained in Ref. \cite{FKS84}, in the case of
mixing of close {\sl single-particle} states one should not expect
necessarily a large enhancement. For example, in a simple
approximate model, where the strong nuclear potential is
proportional to nuclear density and the spin-orbit interaction is
neglected, the matrix element $\langle {\rm g.s.}|W|x\rangle$
contains the single-particle momentum operator. This matrix
element is proportional to $(E_{{\rm g.s.}}-E_{x})$, so that the
small energy denominator cancels out. As mentioned above, the {\sl
collective} Schiff moments in nuclei with static octupole
deformation may be by 2-3 orders of magnitude stronger than
single-particle moments in spherical nuclei.

Consider the following example of a mechanism generating a
collective Schiff moment in spherical nuclei \cite{FZ03}. Let an
odd-$A$ nucleus have the unperturbed ground state of spin $J$
built as a zero spin core plus an unpaired nucleon in the
spherical mean field orbit with angular momentum $j=J$. The
interaction between the odd particle and vibrations of the core
causes an admixture of a quadrupole phonon to the ground state if
the nuclear spin $J> 1/2$:
\begin{equation}
|{\rm g.s.}\rangle=a_{0}|j=J\rangle+a_{2}|[j\otimes
2^{+}]_J\rangle.                                 \label{2.2}
\end{equation}
An opposite parity state with the same spin $J$ can be formed by
coupling an octupole phonon to the ground state particle,
\begin{equation}
|x\rangle=|[j\otimes 3^{-}]_J\rangle.          \label{2.3}
\end{equation}
A (${\cal P},{\cal T}$)-odd Schiff moment (\ref{2.1}) can mix the
states $|{\rm g.s.}\rangle$ and $|x\rangle$. To first order in the
non-relativistic nucleon velocity $p/m$, the (${\cal P},{\cal
T}$)-odd interaction can be presented as \cite{FKS84}
\begin{equation}
W_{ab}=\frac{G}{\sqrt{2}}\frac{1}{2m}\Bigl( (\eta_{ab}
\mbox{\boldmath$\sigma$}_{a}-\eta_{ba}\mbox{\boldmath$\sigma$}_{b})\cdot
\mbox{\boldmath$\nabla$}_{a}\delta({\bf r}_{a}-{\bf r}_{b})+
\eta'_{ab}\left[ \mbox{\boldmath$\sigma$}_{a}\times
\mbox{\boldmath$\sigma$}_{b}\right] \cdot \left\{({\bf p}_{a}-{\bf
p}_{b}),\delta({\bf r}_{a}-{\bf r}_{b}) \right\}\Bigr),
                                                   \label{2.4}
\end{equation}
where $\{\ ,\ \}$ is an anticommutator, $G$ is the Fermi constant
of the weak interaction, $m$ is the nucleon mass, and
\mbox{\boldmath$\sigma$}$_{a,b}$, ${\bf r}_{a,b}$, and ${\bf
p}_{a,b}$ are the spins, coordinates, and momenta, respectively,
of the interacting nucleons $a$ and $b$. The dimensionless
constants $\eta _{ab}$ and $\eta '_{ab}$ characterize the strength
of the (${\cal P},{\cal T}$)-odd nuclear forces; in fact,
experiments on measurement of the EDMs are aimed at extracting the
values of these constants.

In the case of {\sl static} deformation, in the ``frozen" {\sl
body-fixed} frame the {\sl intrinsic} collective Schiff moment
$S_{\rm intr}$ of the deformed nucleus can exist without any
(${\cal P},{\cal T}$)-violation. The estimate found in Refs.
\cite{AFS96,SAF97} gives
\begin{equation}
S_{\rm intr} \approx \frac{9}{20\pi \sqrt{35}}\,eZR^{3}\beta
_{2}\beta_{3}=\frac{3}{5\sqrt{35}}O_{\rm intr}\beta_{2} \ ,
                                         \label{2.5}
\end{equation}
where  $\beta_{2}$ and $\beta_{3}$ are the static quadrupole and
octupole deformation parameters, respectively, and $O_{\rm intr}$
is the static octupole moment. Of course, in the {\sl space-fixed}
laboratory frame, the nucleus has definite angular momentum rather
than fixed orientation, and this makes the expectation value of
the Schiff moment to vanish in the case of no (${\cal
P,T}$)-violation.

The relation (\ref{2.5}) is expected to hold \cite{FZ03} for the
{\sl dynamic} quadrupole and octupole deformations in systems with
spherical equilibrium shape if the effective dynamic deformation
parameters $\beta_{2}$ and $\beta_{3}$ (measured by the multipole
transition strength for the phonon states in the adjacent
even-even nucleus) have a magnitude similar to that in deformed
nuclei. This can be the case under an assumption that the ground
state of the considered odd-$A$ nucleus can be well approximated
by an unpaired particle with an admixture of a single phonon of a
corresponding multipolarity. It is therefore important to try to
perform a detailed calculation and learn more about the validity
of such estimates. Below we present such calculations in the
framework of the conventional QRPA for quadrupole and octupole
phonons.

\section{QRPA phonons and particle-phonon coupling}

We start with the calculation of quadrupole ($J^\pi =2^+$) and
octupole ($J^\pi =3^-$) phonon states in heavy even-even nuclei
based on a conventional model Hamiltonian,
\begin{equation}
H=H_{\rm s-p} + H_{\rm pair} + H_{\rm phon}.    \label{3.1}
\end{equation}
The first term, $H_{\rm s-p}$, describes the single-particle mean
field. The following Woods-Saxon parameterization \cite{dudek81}
was used in calculations:
$$V_c^{p(n)}=-49.6\left(1+(-)0.86\,{N-Z \over A} \right) {\rm MeV}$$
and $$V_{ls}=18.8\left({A \over A-1}\right)^2 {\rm MeV}$$ are the
strength parameters for the central and spin-orbital potentials,
respectively; $R=1.3\,A^{1/3}$ fm and $a=0.7$ fm are the radius
and diffuseness parameters.   The second term of eq. (\ref{3.1}),
$H_{\rm pair}$, is the pairing interaction with the strength
$$G_p ={17.9 +0.176\,(N-Z) \over A}$$ for protons and
$$G_n = {18.95 -0.078\,(N-Z) \over A}$$
for neutrons. The BCS formalism was used that yields a
quasiparticle basis and corresponding Bogoliubov transformation
coefficients $u_j$ and $v_j$.  The last term,
\begin{equation}
H_{\rm phon}=\sum_{\lambda \mu}\omega_\lambda Q^{\dagger}_{\lambda
\mu} Q_{\lambda \mu},                               \label{3.2}
\end{equation}
presents RPA phonons obtained with a separable multipole-multipole
interaction. The building blocks of the model are the
two-quasiparticle RPA phonons,
\begin{equation}
Q^\dagger_{\lambda\mu}=\frac{1}{2} \sum_{j_1,j_2}\left[A_{j_{1}
j_{2}}^{\lambda}
[\alpha^\dagger_{j_1}\alpha^\dagger_{j_2}]_{\lambda\mu}
-(-)^{\lambda-\mu} B_{j_{1}j_{2}}^{\lambda}
[\alpha_{j_1}\alpha_{j_2}]_{\lambda\,-\mu} \right], \label{3.3}
\end{equation}
where $A_{j_1 j_2}$ and $B_{j_1 j_2}$ are the forward and backward
phonon amplitudes,
\begin{equation}
A_{j_1 j_2}^\lambda = \frac{1}{\sqrt{2Z(\lambda)}}\,
\frac{f^\lambda(j_1j_2)\xi^{(+)}_{j_1 j_2}}{
\varepsilon(j_1,j_2)-\omega_{\lambda}}; \quad B_{j_1 j_2}^\lambda
= \frac{1}{\sqrt{2Z(\lambda)}}\,
\frac{f^\lambda(j_1j_2)\xi^{(+)}_{j_1 j_2}}{ \varepsilon(j_1,j_2)+
\omega_{\lambda}},                             \label{3.4}
\end{equation}
where $Z(\lambda)$ is a normalization factor, $f^\lambda(j_1j_2)=
<j_1||r^\lambda Y_\lambda||j_2>$, and the coherence
factors of the Bogoliubov canonical transformation are
$\xi^{(\pm)}_{12}= u_1v_2 \pm v_1u_2$ and
$\eta^{(\pm)}_{12}=u_1u_2 \pm v_1v_2$.

The phonon frequencies $\omega_{\lambda}$ are the roots of the
characteristic RPA equation for each multipolarity $\lambda$,
\begin{equation}
X(\lambda) \equiv \frac{1}{2\lambda+1} \sum_{j_1,j_2}
\frac{[f^\lambda(j_1j_2)\xi^{(+)}_{j_1 j_2}]^2
\varepsilon(j_1,j_2)}{ \varepsilon^2(j_1,j_2)-\omega_{\lambda}^2}
=\frac{1}{\chi(\lambda)},                       \label{3.5}
\end{equation}
where $\chi(\lambda)$ is the strength parameter and
$\varepsilon(j_1,j_2)=\varepsilon(j_1)+ \varepsilon(j_2)$ is the
unperturbed two-quasiparticle energy. Solving these equations one
obtains the energies of the phonons and internal structure of the
phonon operator (\ref{3.3}) hidden in the amplitudes of different
two-quasiparticle components. The values of the strength
parameters, $\chi(2)$ and $\chi(3)$, were chosen to reproduce the
excitation energies of the $2^+_1$ and $3^-_1$ states in the
neighboring even-even nuclei.

We are interested in the evaluation of the Schiff moment for the
odd-neutron nuclei. Therefore the second step of calculations is
the solution for the odd-$A$ nucleus with the even-even core
excitations described above. If we neglect the
(quasiparticle+two-phonon) components in the wave function of an
excited state of an odd-$A$ nucleus, then the corresponding wave
functions have the following form:
\begin{equation}
\Psi_n(J^\pi)=C_{Jn}\Bigl(\alpha_{JM}^{\dagger}\delta_{jJ} +
\sum_{\lambda,j}D_{Jj}^{\lambda,n} [\alpha_{j}^{\dagger}\otimes
Q^+(\lambda)]_{JM}\Bigr)\Psi_0,                 \label{3.6}
\end{equation}
where the operator $\alpha^{\dagger}$ creates a quasiparticle with
respect to the ground state wave function $\Psi_{0}$ of the
even-even nucleus. The energy $E_n(J)$ of the $n^{{\rm th}}$ state
with angular momentum $J$ in the odd-mass nucleus, the amplitudes
of the quasiparticle-phonon components,
\begin{equation}
D_{Jj}^{\lambda,n}=\sqrt{\frac{2\lambda+1}{(2J+1)2Z(\lambda)}}\,
\frac{f^\lambda(Jj)\eta^{(-)}_{J,j}}{
\varepsilon(j)+\omega_{\lambda}-E_n},            \label{3.7}
\end{equation}
and the amplitude of the single-quasiparticle component,
\begin{equation}
C_{Jn}^{-2}=1+\sum_{\lambda, j} (D_{Jj}^{\lambda,n})^2,
                                                 \label{3.8}
\end{equation}
are obtained from the solution of the secular equation
\cite{Sol92}:
\begin{equation}
\varepsilon(J) - E_n =\sum_{\lambda,j}
\left(D_{Jj}^{\lambda,n}\right)^2 \left[
\varepsilon(j)+\omega_{\lambda}-E_n\right].    \label{3.9}
\end{equation}

We performed the calculations for several isotopes of radium and
radon. The lighter even-even isotopes, $A=216-218$, are considered
to be still spherical; the ones with $A=220$ are transitional and
the heavier isotopes $A=222-224$ acquire more pronounced deformed
structure. In the six even-even nuclei treated here, see Table 1,
the $2^{+}$ and $3^{-}$ states were experimentally observed with
excitation energies $\omega_{2}$ and $\omega_{3}$ listed in Table
1. The values of the coupling constants $\chi(2)$ and $\chi(3)$,
also shown in this Table, were found by fitting the lowest RPA
$2^{+}$ and $3^{-}$ states to the corresponding experimental
excitation energies.

\begin{table}
\label{T1} \caption{Phonon energies in even-even nuclei and the
corresponding parameters of the multipole-multipole interaction.}
\begin{center}
\begin{tabular}{c|c|c|c|c}
\hline
 Nucleus & $\omega_{2}$, MeV & $\omega_{3}$, MeV &
 $\chi(2),\,10^{-4}$ & $\chi(3),\,10^{-5}$ \\
 \hline
 $^{216}_{88}$Ra$_{128}$ & 0.69 & 1.40 & 8.1 & 1.2 \\
 $^{216}_{86}$Rn$_{130}$ & 0.46 & 1.11 & 6.9 & 1.2 \\
 $^{218}_{88}$Ra$_{130}$ & 0.39 & 0.79 & 6.9 & 1.2 \\
 $^{218}_{86}$Rn$_{132}$ & 0.32 & 0.84 & 5.9 & 1.1 \\
 $^{220}_{88}$Ra$_{132}$ & 0.18 & 0.47 & 5.9 & 1.1 \\
 $^{220}_{86}$Rn$_{134}$ & 0.24 & 0.66 & 5.3 & 1.1

 \end{tabular}
 \end{center}
 \end{table}

The single-particle states were calculated in a spherical
Woods-Saxon potential with the parameters indicated above. For
heavier isotopes this might not be an appropriate approximation;
the purpose of our calculations for these nuclei was to examine
the systematics of the results in the cases when the phonons are
lowered in energy.

\begin{table}
\label{T2} \caption{Structure of the ground ($J^{\pi}$) and
excited state with the same spin and opposite parity ($J^{-\pi}$)
in odd-neutron nuclei (probabilities of various components in the
wave functions).}
\begin{center}
\begin{tabular}{c|c|cc|c|cc}
\hline
 $^{217}_{88}$Ra$_{129}$ & $9/2^{+}$ & $2g_{9/2}$ & 0.71 &
$9/2^{-}$ & $1h_{9/2}$ & $10^{-4}$ \\
& & $2g_{9/2}\otimes 2^{+}$ & 0.14 & & $2g_{9/2}\otimes 3^{-}$ &
0.999 \\
& & $1j_{15/2}\otimes 3^{-}$ & 0.08 & & & \\
& & $3d_{5/2}\otimes 2^{+}$ & 0.05 & & & \\
\hline
 $^{217}$Rn$_{131}$ & $9/2^{+}$ & $2g_{9/2}$ & 0.73 &
$9/2^{-}$ & $1h_{9/2}$ & $2\cdot 10^{-4}$ \\
& & $2g_{9/2}\otimes 2^{+}$ & 0.11 & & $2g_{9/2}\otimes 3^{-}$ &
0.999 \\
& & $1j_{15/2}\otimes 3^{-}$ & 0.07 & & & \\
& & $3d_{5/2}\otimes 2^{+}$ & 0.07 & & & \\
\hline
 $^{219}_{88}$Ra$_{131}$ & $7/2^{+}$ & $2g_{7/2}$ & 0.43 &
$7/2^{-}$ & $2f_{7/2}$ & 0.02 \\
& & $1i_{11/2}\otimes 2^{+}$ & 0.27 & & $2g_{9/2}\otimes 3^{-}$ &
0.98 \\
& & $2g_{7/2}\otimes 2^{+}$ & 0.11 & & & \\
& & $2g_{9/2}\otimes 2^{+}$ & 0.08 & & & \\
& & $3d_{3/2}\otimes 2^{+}$ & 0.09 & & & \\
\hline
 $^{219}_{86}$Rn$_{133}$ & $5/2^{+}$ & $3d_{5/2}$ & 0.48 &
$5/2^{-}$ & $ 2f_{7/2}$ & 0.01 \\
& & $2g_{9/2}\otimes 2^{+}$ & 0.27 & & $ 2g_{9/2}\otimes 3^{-}$ &
0.98 \\
& & $3d_{5/2}\otimes 2^{+}$ & 0.12 & & & \\
& & $4s_{1/2}\otimes 2^{+}$ & 0.09 & & & \\
\hline
 $^{221}_{88}$Ra$_{133}$ & $5/2^{+}$ & $3d_{5/2}$ & 0.48 &
$5/2^{-}$ & $2f_{5/2}$ & 0.04 \\
& & $2g_{9/2}\otimes 2^{+}$ & 0.25 & & $2g_{9/2}\otimes 3^{-}$ &
0.86 \\
& & $3d_{5/2}\otimes 2^{+}$ & 0.13 & & $1i_{11/2}\otimes 3^{-}$ &
0.09 \\
& & $4s_{1/2}\otimes 2^{+}$ & 0.10 & & & \\
& & $3d_{3/2}\otimes 2^{+}$ & 0.03 & & & \\
\hline
 $^{221}_{86}$Rn$_{135}$ & $7/2^{+}$ & $2g_{7/2}$ & 0.44 &
$7/2^{-}$ & $2f_{7/2}$ & 0.02 \\
& & $1i_{11/2}\otimes 2^{+}$ & 0.26 & & $2g_{9/2}\otimes 3^{-}$ &
0.98\\
& & $2g_{7/2}\otimes 2^{+}$ & 0.12 & & & \\
& & $3d_{3/2}\otimes 2^{+}$ & 0.12 & & & \\
& & $2g_{9/2}\otimes 2^{+}$ & 0.04 & & &

\end{tabular}
\end{center}
\end{table}

In Table II we show the composition of the ground state wave
functions and their suitable parity partners in the six
odd-neutron nuclei considered in this work. One can see from the
wave functions that while the dominant component for negative
parity is the $3^{-}\otimes g_{9/2}$ configuration, the
corresponding component $2^{+}\otimes g_{9/2}$ is of the order of
10-30\%. The spacings between the ground states and their parity
partners turn out to be of order 1-2 MeV, Table I. The mixing of
these two components will contribute the most to the Schiff
moment. We should note here that the calculated spectra of the
odd-neutron nuclei do not agree well with experiment even in the
cases of $A=217,\, 219$. This indicates that the model used for
the description of coupling of the odd particle with the core
vibrations is too restrictive.

\section{The Schiff moment}

Having determined the wave functions one may proceed with the
calculation of the Schiff moment according to Eq. (\ref{2.1}),
where we will use the notations $|J_g^\pi\rangle$ and
$|J_g^{-\pi}\rangle$ for the ground state of the odd-neutron
nucleus and for the first excited state of opposite parity,
respectively,
\begin{equation}
S(J_g^\pi)=2\, \frac{ \langle J_g^\pi | W | J_g^{-\pi}  \rangle
\langle J_g^{-\pi}J_g | S_z | J_g^\pi J_g\rangle}{E (J_g^\pi) - E
(J_g^{-\pi} )}.                                  \label{4.1}
\end{equation}
Here $W$ is the (${\cal P,T}$)-violating nucleon-nucleon
interaction given by Eq. (\ref{2.4}) and $\langle J_g^{-\pi}J_g| S
| J_g^\pi J_g\rangle$ is the matrix element of the Schiff operator
${\bf S}$,
\begin{equation}
S_\mu=\frac{1}{10}\sqrt{4\pi \over 3}\,
\sum_iY_{1\mu}(\theta_i,\varphi_i)e_{i}\left[ r_i^{3}
-\frac{5}{3}\, r^{2}_{\rm ch}r_i \right],     \label{4.2}
\end{equation}
for the maximum projection $M=J_{g}$ of angular momentum, and
$r^{2}_{\rm ch}$ is the mean square charge radius.

Since only the proton components of the phonons contribute to the
matrix element of the Schiff moment, this matrix element for the
case of the odd neutron can be written as
\[\langle J^\pm_g M=J_g| S_z | J^\mp_g M=J_g \rangle\]
\begin{equation}
= \sqrt{J_g(2J_g+1) \over (J_g+1)} \sum_{j,\lambda,\lambda'}
 (-1)^{j+\lambda'+J_g+1}
\tilde{D}^\lambda_{J_g^\pm,j}\tilde{D}^{\lambda'}_{J_g^\mp,j} \left\{
\begin{array}{rrr}
  \lambda & \lambda' &  1    \\
  J_g & J_g & j \\
  \end{array}
\right\} \langle \lambda \| S \| \lambda' \rangle, \label{4.3}
\end{equation}
where $\tilde{D}= C\cdot D$ - is the amplitude of the
(quasiparticle plus phonon) component, and the sum over $\lambda$
and $\lambda'$ is in reality reduced to one term with
$\lambda^\pi=2^+$ and $\lambda'^\pi=3^-$. The matrix element of
the Schiff operator between the quadrupole and octupole phonon
states in the even nucleus has the following form in terms of the
RPA amplitudes (\ref{3.4}):
\begin{equation}
\langle 2^+||S ||3^- \rangle =
\sqrt{35}\sum_{j_1,j_2,j_3}\eta^{(-)}_{j_1,j_2} \langle j_1 \|S
\|j_2 \rangle \left\{
\begin{array}{rrr}
  2 & 3 &  1    \\
  j_1 & j_2 & j_3 \\
  \end{array}
\right\} (A^{(2+)}_{j_2 j_3}A^{(3-)}_{j_3 j_1} + B^{(2+)}_{j_2
j_3}B^{(3-)}_{j_3 j_1}).                       \label{4.4}
\end{equation}
The results of the calculations are shown in Table III.

\section{Matrix element of weak interaction}

The calculation of the weak interaction matrix element appearing
in Eqs. (\ref{2.1}) and (\ref{4.1}) is elaborate and contains a
number of contributions.  We collect these calculations more in
detail in Appendix. Here we present only the final expressions for
the important contributions and numerical results, Table III.

\begin{table}
\label{T3} \caption{Mixing matrix elements of weak interaction
(m.e. $W$), of the Schiff moment (m.e. $S$), and final value for
the Schiff moment, $S$, in the ground state of the odd-neutron nucleus.}
\begin{center}
\begin{tabular}{c|c|c|c|c|c}
\hline
 Nucleus & $E_{+}$, MeV & $E_{-}$, MeV & m.e. $W$, $\eta\cdot
 10^{-2}$ eV & m.e. $S$, $e\cdot$fm$^{3}$ & $S$, $\eta\cdot 10^{-8}e\cdot$fm$^{3}$ \\
\hline
 $^{217}_{88}$Ra$_{129}$ & 0.09 & 2.44 & 0.3 & -0.1 & $-0.03$ \\
\hline
 $^{217}_{86}$Rn$_{131}$ & $-0.12$ & 1.80& $0.1$ & -0.1 &$-0.01$ \\
\hline
 $^{219}_{88}$Ra$_{131}$ & 0.49 & 1.47& -1.3 & -0.1 & 0.30 \\
\hline
 $^{219}_{86}$Rn$_{133}$ & -0.09 & 1.42 & $0.2$ & -0.1 & $-0.03$ \\
\hline
 $^{221}_{88}$Ra$_{133}$ & $-0.77$ & 1.05 & $0.2$ & -0.2 & $-0.07$\\
\hline
 $^{221}_{88}$Rn$_{135}$ & 0.24 & 1.16 & -0.5 & -0.1 & 0.06

\end{tabular}
\end{center}
\end{table}

In this work we account only for the direct contribution of the
weak interaction, Eq. (\ref{2.4}), to the nuclear mean field for a
particle $b=p,n$ at a point ${\bf r}=r{\bf n}$,
\begin{equation}
\overline{W}_b({\bf r}) = \frac{G}{\sqrt{2}}\frac{1}{2m} \sum_c
\eta_{bc} (\mbox{\boldmath$\sigma$} {\bf n}) \frac{d}{dr} \frac{2
J_c +1}{4 \pi} v^2_c R^2_{c}(r) = \frac{G}{\sqrt{2}}\frac{1}{2m}
\eta \, (\mbox{\boldmath$\sigma$} {\bf n}) \frac{1}{4\pi}\frac{d
\rho(r)}{dr},                              \label{5.1}
\end{equation}
where $v_{c}^{2}$ is the occupancy of the single-particle state
$c$ in the core. We expect this term to be the largest since all
particles in the core add coherently in Eq. (\ref{5.1}). The weak
matrix elements are typically of the order $10^{-2}\,\eta$ eV. The
main contribution comes from the mixing between the (quasiparticle
plus phonon) component of the excited state and the pure
quasiparticle component of the ground state.

The reduced matrix elements $\langle 2^{+}||S||3^{-}\rangle$ in
even nuclei are of the order (1-2)$e\cdot{\rm fm}^{3}$, and the
matrix elements of the Schiff operator between the ground state
and its parity partner turn out to be around $0.1-0.2$ in units of
$e\cdot{\rm fm}^{3}$. The results for the Schiff moment in this
model are within the limits (0.04-0.6)$\eta\cdot
10^{-8}\,e\cdot{\rm fm}^{3}$, i.e. of the same order as in pure
single-particle models in even nuclei. The single-particle
contribution unrelated to the soft modes \cite{FKS86,DS03,DS04} is
of the same order of magnitude and it should be added.

\section{Discussion}

How can one reconcile this result with the large enhancement found
in the calculations for the nuclei with static quadrupole+octupole
deformation? We need to stress that in those calculations the
deformations are introduced explicitly from the beginning (in some
calculations, only the quadrupole deformation is taken at a
starting point while the octupole part is introduced via coupling
of a particle to the $3^{-}$ phonon of the deformed core).
Apparently, it is essential to start from a particle moving in a
quadrupole deformed field in order to find large enhancement
factors for the Schiff moment. The attempt to reproduce this
effect by having in the ground state of the odd-$A$ nucleus an
admixture of a particle plus a single $2^{+}$ phonon can succeed
only in limiting cases.

In the case of static quadrupole, $\beta_{2}$, and octupole,
$\beta_{3}$, deformations, the weak matrix element is
$\propto\beta_{3}$, the Schiff moment operator $\propto
\beta_{2}\beta_{3}/\Delta E_{\pm}$ as discussed in
\cite{AFS96,SAF97}. Although the small denominator $\Delta
E_{\pm}\sim 50$ keV was the main source of enhancement, it was
also important to have noticeable values of the deformation
parameters. Since in dynamical models with low phonon frequency
$\omega$ the dynamic deformation parameters $\beta\propto
1/\omega$, one may expect a considerable enhancement at
$\omega\rightarrow 0$, close to the phase transitions to static
deformation. Such limiting cases will be discussed below.

The effects of quadrupole deformation can be mimicked by a high
degree of coherence in a quadrupole mode. Similarly to that, the
transition of a spherical nucleus to an octupole deformed state
was discussed in the RPA framework in the past \cite{abbas81}. It
was found that even for realistic interactions the energies of the
lowest $3^{-}$ RPA root collapse in certain areas of the periodic
table. The stability of the nucleus against variation of a certain
collective variable $Q$ can be expressed in terms of the static
polarizability defined for instance through the dependence of
constrained Hartree-Fock energy $E/A$ on the constraining term
$\alpha Q$,
\begin{equation}
P=\frac{1}{2}\,\left[\frac{\partial^{2}(E/A)}{\partial \alpha^{2}}
\right]_{\alpha=0}.                           \label{7.1}
\end{equation}
In the RPA, this quantity can be calculated as
\begin{equation}
P=2\,\frac{m_{-1}}{A}                        \label{7.2}
\end{equation}
by relating it to the $m_{-1}$ moment (the inverse energy weighted
sum rule),
\begin{equation}
m_{-1}=\sum_{n}\,\frac{|\langle n|Q|0\rangle|^{2}}{\omega_{n}},
                                          \label{7.3}
\end{equation}
with $\omega_{n}$ being the RPA energies corresponding to
excitation from the ground state $|0\rangle$ to the excited state
$|n\rangle$ with quantum numbers dictated by the symmetry of the
operator $Q$. As polarizability increases the system becomes less
stable against deformation characterized by the collective
variable $Q$; a negative value of $P$ indicates a situation in
which a collapse has occurred (in the RPA framework).

In order to take advantage of such properties and examine the
behavior of the Schiff moment in situations close to the RPA
collapse that signals a transition from a spherical to a
pear-shaped nucleus (quadrupole + octupole deformation), we can
consider the limit of the RPA frequencies $\omega_{2}$ and
$\omega_{3}$ going to zero. Introducing formally a small scaling
factor $y$ ,
\begin{equation}
\omega_{2,3}\rightarrow y\omega_{2,3}, \quad  y\ll 1, \label{7.4}
\end{equation}
we separate the singular in this limit part of the secular
equation (\ref{3.5}) as
\begin{equation}
X(\lambda)=\frac{1}{2\lambda+1}\,\sum_{12} \frac{|f^{\lambda}_{12}
\xi^{(+)}_{12}|^{2}}{\epsilon_{1}+\epsilon_{2}}+O(y^{2}).
                                                    \label{7.5}
\end{equation}
The normalization factor $Z(\lambda)$ goes to zero linearly in
$y$, and the phonon amplitudes (\ref{3.4}) grow $\propto
1/\sqrt{y}$. As a result of this overcritical coupling, the energy
of the ground state in the odd nucleus collapses as well (goes to
$-\infty$ as $1/\sqrt{y}$). The wave function components
(\ref{3.6}) have a constant limit with the single-particle
amplitude $C^{2}\rightarrow 1/2$; the same effect was observed in
\cite{StZ04} for the strong coupling of a quasiparticle with
phonons.

\begin{table}
\label{T4} \caption{Scaling of RPA solutions with the low-lying
frequency of collective oscillations in the even core:
normalization factors $Z(\lambda)$; the reduced matrix element
$\langle 2||S||3\rangle$ of the Schiff moment in the even nucleus;
energies $E_{+}$ and $E_{-}$ of the ground and excited state in
the odd nucleus, in MeV; the matrix element (m.e. $W$) of the weak
interaction between these states; the matrix element (m.e. $S$) of
the Schiff moment in the odd nucleus; and the final result for the
expectation value of the Schiff moment; the matrix elements of $W$
and $S$ are given in the same units as in Table III. }

\begin{center}
\begin{tabular}{c|c|c|c|c|c|c|c|c|c}
\hline
 Nucleus & $y$ & $Z(2)$ & $Z(3)$ & $\langle 2||S||3\rangle$ &
$E_{+}$ & $E_{-}$ & m.e. $W$ & m.e. $S$ & S \\
\hline
 $^{219}$Ra & 1.0 & 479 & 21696 & 1.7 & 0.5 & 1.5 & $-1.3$ & $-0.1$ &
0.3 \\
 & 0.1 & 43 & 1825 & 20 & $-3.8$ & $-1.9$ & 1.1 & $-0.2$ & $-0.2$ \\
 & 0.01 & 4 & 182 & 195 & $-17$ & $-19$ & 53 & $-0.2$ & 6.2 \\
\hline
\hline
 $^{221}$Ra & 1.0 & 324 & 18100 & 2.2 & $-0.8$ & 1.0 & 0.2 & $-0.2$ &
$-0.1$ \\
& 0.1 & 31 & 1650 & 23 & $-6.1$ & $-2.9$ & $-19$ & $-0.5$ & 6 \\
& 0.01 & 3 & 165 & 235 & $-23$ & $-21$ & $-253$ & $-2.7$ & 560

\end{tabular}
\end{center}
\end{table}

In Table IV the results of the calculation for $^{219,221}$Ra are
shown as a function of $y$-scaling. As the collective frequencies
go down, the reduced matrix element $\langle 2||S||3\rangle$ in
the even nucleus, the matrix element of the weak interaction in
the odd nucleus and the resulting Schiff moment become large. This
is especially pronounced in $^{221}$Ra, where the Schiff moment is
strongly enhanced at the collapse threshold.

In the RPA framework the matrix element of the Schiff moment
between the states of opposite parity in the odd nucleus can be
presented in terms of the amplitudes $\tilde{D}^{\lambda}_{jj'}$, eq.
(\ref{3.6}), as
\begin{equation}
\langle +|S|-\rangle\propto\sum_{\lambda_{1}\lambda_{2}j'}
\tilde{D}^{\lambda_{1}(+)}_{jj'} \tilde{D}^{\lambda_{2}(-)}_{jj'}
\langle\lambda_{1}||S||\lambda_{2}\rangle.     \label{7.6}
\end{equation}
In the case of $^{219}$Ra, at the experimental values of the
phonon frequencies, $y=1$, there exists only one large combination
of amplitudes,
\begin{equation}
|\tilde{D}^{2(+)}_{g_{9/2},g_{9/2}}|^{2}=0.08, \quad
|\tilde{D}^{3(-)}_{g_{9/2},g_{9/2}}|^{2}=0.98.         \label{7.7}
\end{equation}
Here the excited negative parity state is an almost pure
combination $2g_{9/2}\otimes 3^{-}$. With reduction of
frequencies, the amplitudes (\ref{7.7}) rapidly decrease going to
limiting values of $|\tilde{D}^{2(+)}|^2\approx 0.01$ and
$|\tilde{D}^{3(-)}|^2\approx 0.02$. The wave function
(quasiparticle + phonon) turns out to be spread over many states,
and the enhancement of the Schiff moment due to the growth of the
matrix element $\langle 2||S||3\rangle$ starts with further
diminishing frequencies only after the spreading saturates in a
given single-particle space. The natural conclusion is that the
enhancement of the Schiff moment in the presence of soft
quadrupole and octupole modes is a real physical effect which is
however hindered by the complexity of the mixed states in the
limit of strong coupling between quasiparticles and soft mode
phonons.

However, the ansatz (\ref{3.6}) for the wave function of the odd
nucleus becomes invalid in this limit, and many-phonon components
carry a large fraction of the total normalization. Leaving the
full solution of this problem for the future, we just show that
the enhancement occurs beyond the RPA framework.

\section{Simple model beyond RPA}

The strong quasiparticle-phonon coupling leads to stationary
states which are close to the coherent states, or condensates, of
the phonon field \cite{BZ65}. This limit is different from that
used in (\ref{3.6}), where only a one-phonon admixture was taken
into account.

We consider the Hamiltonian (for simplicity with one
single-particle $j$-level but the extension is straightforward):
\[H=\epsilon\sum_{m}\alpha^{\dagger}_{m}\alpha_{m}+\sum_{\lambda\mu}
\omega_{\lambda}Q^{\dagger}_{\lambda\mu}Q_{\lambda\mu}\],
\begin{equation}
+\sum_{\lambda\mu; mm'}x_{\lambda}\left(\begin{array}{ccc}
 \lambda & j & j \\
 \mu & -m & m'\end{array}\right)(-)^{j+m}
 \hat{\rho}_{m'm}\Bigl(Q_{\lambda\mu}+(-)^{\lambda-\mu}
 Q^{\dagger}_{\lambda -\mu}\Bigr),              \label{8.1}
\end{equation}
where $x_{\lambda}$ are the parameters of the particle-phonon
coupling, $Q_{\lambda\mu}$ phonon operators, and
$\hat{\rho}_{m'm}=a^{\dagger}_{m}a_{m'}$ is the operator of the
single-particle density matrix that satisfies the operator
equation of motion:
\begin{eqnarray}
[\hat{\rho}_{m'm},H]=\sum_{\lambda\mu; m''}x_{\lambda}
\left(\begin{array}{ccc}
 \lambda & j & j \\
 \mu & - m'& m''\end{array}\right)(-)^{j+m'}\hat{\rho}_{m''m}
 \Bigl(Q_{\lambda\mu}+(-)^{\lambda-\mu}
 Q^{\dagger}_{\lambda -\mu}\Bigr) \\ \nonumber
 -\sum_{\lambda\mu; m''}x_{\lambda}\left(\begin{array}{ccc}
 \lambda & j & j \\
 \mu & -m& m''\end{array}\right)(-)^{j+m}\hat{\rho}_{m'm''}
 \Bigl(Q_{\lambda\mu}^{\dagger}+(-)^{\lambda-\mu}
 Q_{\lambda -\mu}\Bigr).                     \label{8.2}
\end{eqnarray}
The phonon operators satisfy
\begin{equation}
[Q_{\lambda\mu},H]=\omega_{\lambda}Q_{\lambda\mu}+x_{\lambda}
\sum_{mm'}(-)^{\lambda-\mu}\left(\begin{array}{ccc}
 \lambda & j & j \\
- \mu & -m & m'\end{array}\right)(-)^{j+m}
 \hat{\rho}_{m'm}.                           \label{8.3}
\end{equation}

For monopole phonons, $\lambda=0$, the exact solution \cite{BZ65}
leads to the coherent cloud of phonons with the Poisson
distribution of the phonon numbers. For $\lambda\neq 0$ the
analytical solution does not exist; however, we can use the method
of Ref. \cite{BZ70} that allows to separate the main effects in
the adiabatic limit of slow collective motion. We assume that the
even core is spherical with low-frequency collective modes,
$\omega_{2}$ and $\omega_{3}$. Then in the odd nucleus we have a
phonon condensate $\langle Q_{\lambda\mu}\rangle$ that can be
found from (\ref{8.3}),
\begin{equation}
\langle Q_{\lambda\mu}\rangle=-\frac{x_{\lambda}}
{\omega_{\lambda}} \sum_{mm'}(-)^{\lambda-\mu}
\left(\begin{array}{ccc}
 \lambda & j & j \\
- \mu & -m & m'\end{array}\right)(-)^{j+m}
 \langle \rho_{m'm}\rangle.                    \label{8.4}
\end{equation}
This condensate is created by the average odd-particle density
matrix $\langle \rho_{m'm}\rangle$ that determines
self-consistently a direction of the intrinsic quantization axis.
This direction is arbitrary in agreement with full rotational
invariance, and we can expand this density matrix over multipoles,
\begin{equation}
\langle \rho_{m'm}\rangle=\sum_{L\Lambda}(-)^{L+j-m}
\left(\begin{array}{ccc}
 L & j & j \\
 \Lambda & -m & m' \end{array}\right)
 Y_{L\Lambda}\rho_{L},                         \label{8.5}
\end{equation}
where the spherical function shows the ``rotational behavior" of
the condensate in the laboratory frame of the odd nucleus. Then we
find the relation between the condensate and the density matrix
multipoles through effective deformation parameters
$\beta_{\lambda}$,
\begin{equation}
\langle
Q_{\lambda\mu}\rangle=\beta_{\lambda}Y^{\ast}_{\lambda\mu}, \quad
\beta_{\lambda}=\frac{1}{\omega_{\lambda}}\,
\frac{x_{\lambda}}{2\lambda+1}\,\rho_{\lambda}.   \label{8.6}
\end{equation}
As mentioned earlier, the effective deformation parameters induced
in the odd nucleus by soft modes of the even core are inversely
proportional to the corresponding frequencies. (In our notations
we have $Y^{\ast}_{\lambda\mu}$ because this function lowers the
projection by $\mu$, in accordance with the definition of
$Q_{\lambda\mu}$.)

The relation between the states of the odd and even nuclei is
established through the equation of motion for single-particle
operators linearized after the substitution of the phonon dynamics
by the condensate,
\begin{equation}
[a_{m},H]=\epsilon a_{m}+2\sum_{\lambda\mu; m'}x_{\lambda}
\left(\begin{array}{ccc}
 \lambda & j & j \\
 \mu & -m & m'\end{array}\right)(-)^{j+m}
 a_{m'}\beta_{\lambda}Y^{\ast}_{\lambda\mu}.         \label{8.7}
\end{equation}
In order to diagonalize this set of equations, we are looking for
the particle operator in the form
\begin{equation}
a_{m}=\sum_{k}D^{j}_{mk}c_{k},                     \label{8.8}
\end{equation}
where $c_{k}$ are particle annihilation operators with projection
$k$ onto the axis defined by the phonons, and $D^{j}_{mk}$ are
Wigner rotational functions.

After simple algebra, this procedure results in the occupation
numbers $n_{k}$ of the particle in the self-consistent phonon
field,
\begin{equation}
\langle c^{\dagger}_{k}c_{k'}\rangle=n_{k}\delta_{kk'},
                                                  \label{8.9}
\end{equation}
and the ``Nilsson-type" energies of the odd particle in the
body-fixed frame,
\begin{equation}
{\cal E}_{k}=\epsilon+\sum_{\lambda}\,\frac{2}
{\sqrt{4\pi(2\lambda+1)}} \,\frac{x_{\lambda}^{2}}
{\omega_{\lambda}} (-)^{j-k+\lambda}
 \left(\begin{array}{ccc}
 \lambda & j & j \\
 0 & k & -k \end{array}\right)\rho_{\lambda},   \label{8.10}
\end{equation}
similar to how it would appear as a result of a self-consistent
mutlipole-multipole interaction. The original lab-frame density
matrix (\ref{8.5}) is diagonalized in the multipole form,
\begin{equation}
\rho_{\lambda}=(-)^{\lambda}\sqrt{\frac{4\pi}{2\lambda+1}}\,
\sum_{k}(-)^{j-k}
\left(\begin{array}{ccc}
 j & j & \lambda \\
 -k & k & 0 \end{array}\right) n_{k}.     \label{8.11}
\end{equation}
In particular,
\begin{equation}
\rho_{0}=-\sqrt{\frac{4\pi}{2j+1}}\,\sum_{k}n_{k}, \label{8.12}
\end{equation}
where we get the total number of unpaired particles (equal to 1 in
our simplified case). The model can be advanced further taking
into account the non-adiabatic corrections to the condensate,
analogously to the frequency expansion in (\ref{7.5}).

Now we can show that there is a strong enhancement of the Schiff
matrix element compared to the value $S^{\circ}\equiv
(2||S_{1}||3)$ in the even nucleus. The collective contribution to
this operator is given by
\begin{equation}
S_{1\nu}=S^{\circ}\sum_{\mu\mu'}(-)^{\nu+\mu'}\left(\begin{array}{ccc}
1 & 2 & 3 \\
-\nu & \mu & -\mu' \end{array}\right)\Bigl(Q^{\dagger}_{3\mu'}
Q_{2\mu}+(-)^{\mu+\mu'}Q^{\dagger}_{2-\mu}Q_{3\mu'}\Bigr).
                                                 \label{8.13}
\end{equation}
Using the deformation parameters (\ref{8.6}) for the odd nucleus,
we obtain
\begin{equation}
\frac{S_{1\nu}}{S^{\circ}}=2\beta_{2}\beta_{3}\sum_{\mu\mu'}
(-)^{\nu+\mu'}\left(\begin{array}{ccc}
 1 & 2 & 3 \\
-\nu & \mu & -\mu' \end{array}\right)Y_{3\mu'}Y_{2-\mu}
=-\,\frac{1}{\sqrt{\pi}}\,\beta_{2} \beta_{3}Y^{\ast}_{1\nu}.
                                            \label{8.14}
\end{equation}
With the self-consistent values of deformation parameters
(\ref{8.6}),
\begin{equation}
\frac{S_{1\nu}}{S^{\circ}}=-\,\frac{1}{35\sqrt{\pi}}\,
\frac{x_{2}x_{3}}{\omega_{2}\omega_{3}}\,\rho_{2}\rho_{3}
Y^{\ast}_{1\nu}.                            \label{8.15}
\end{equation}
Consequently, in the odd nucleus we reduced the problem to that
for intrinsic deformation and corresponding intrinsic Schiff
moment, compare eq. (\ref{2.5}). In contrast to the RPA, the wave
function of the odd particle is not fragmented being a
self-consistent mean field solution in the presence of the
condensate. The enhancement due to small frequencies is clearly
present in (\ref{8.15}) with no reduction that appeared in the RPA
calculation.

\section{Conclusion}

We presented the calculations of the Schiff moment in odd-neutron
nuclei with soft quadrupole and octupole vibrational modes present
in the even-even spherical core. The conventional approach used
above is based on the QRPA which allows one to define
microscopically the structure of collective modes and coupling of
the odd particle to phonons of those modes. We confirmed the
effect of enhancement of the Schiff moment for small quadrupole
and octupole frequencies. The even-even system close to the onset
of deformation acts essentially similarly to the statically
deformed system, where the enhancement was established earlier.

However, the QRPA supplemented by the standard ansatz for the
(particle + one phonon) states in the odd system, turns out to be
inadequate. The enhancement requires strong particle-phonon
coupling. In this case, the single-particle strength is widely
fragmented, and a small weight of mixing configurations (particle
+ quadrupole phonon) and (particle + octupole phonon) with the
same spherical orbit for the particle practically cancels the
enhancement of the Schiff matrix element between $2^{+}$ and
$3^{-}$ phonons. The outcome for the Schiff moment in this type of
calculation is on the same level of magnitude as for purely
single-particle states. The QRPA calculations do not predict small
energy denominators $\Delta_{\pm}$ which would provide a strong
enhancement of the effect. In general, the spectra of odd-neutron
nuclei in the case of soft transitional even core cannot be
reliably reproduced by the standard version of the QRPA. The
experimental (poorly known) spectra of $^{219,221}$Ra show the
states, sometimes with unknown spin, of parity opposite to that of
the ground state at $\Delta E\leq 100$ keV.

We demonstrated the presence of the effect by artificially
diminishing vibrational frequency to the limit when the
single-particle spread is saturated, We also sketched the more
general formalism beyond the RPA that is capable of describing the
strong coupling limit for a particle in the field of soft phonons.
The self-consistently emerging coherent phonon state acts
analogously to static deformation, in agreement with the original
idea. We hope to present corresponding advanced calculations
elsewhere.

We have considered only the lowest symmetric quadrupole phonon in
the present paper. In the presence of few active protons and
neutrons, there is a possibility of low-lying ``mixed-symmetry''
quadrupole phonon in vibrational nuclei that is known to be
connected to the octupole $3^-$ phonon by a considerably stronger
E1 strength than the symmetric lowest quadrupole phonon
\cite{Pietralla}. Therefore one may expect additional enhancement
if the coupling to the ``mixed-symmetry'' phonon is taken into
consideration.

The conventional core polarization effects are also important for
getting a certain quantitative prediction, although they are not
expected to lead to a significant amplification of the effect. In
the future the full calculations taking into account on equal
footing collective and single-particle effects, including the
${\cal P,T}$-violating corrections to the vibrational modes and
core polarization, are to be carried out. It is desirable to
extend the studies to lighter nuclei $A \approx 100$ with
pronounced octupole phonon states and more experimental data
available. In particular, there are indications for the intrinsic
coupling between quadrupole and octupole degrees of freedom, see
global systematics \cite{metlay95} and recent experimental data
\cite{mueller06}. This may help to understand better the
possibility for the enhancement of the Schiff moment and to
explore physical correlations between the calculated
values of the Schiff moment and complex nuclear structure.\\
\\
The work was supported by the NSF grant PHY-0244453 and by the
grant from the Binational Science Foundation US-Israel. V.V.F.
acknowledges support from the Australian Research Council and from
the National Superconducting Cyclotron Laboratory at the Michigan
State University.

\section*{Appendix}

Our wave function of the odd nucleus includes two types of states,
pure quasiparticle (QP) and quasiparticle+phonon. Then, any matrix
element is a sum of four terms. One term mixes pure QP components
of the wave functions, the other two mix QP and QP+phonon
components, and the last one mixes two QP+phonon components. It is
convenient to illustrate graphically different contributions to
the weak matrix element.

\subsubsection{QP and QP Mixing}
Here we calculate the weak matrix element between pure QP
components in the mean field approximation (\ref{5.1}):
\begin{figure}
\begin{center}
\includegraphics[width=12cm]{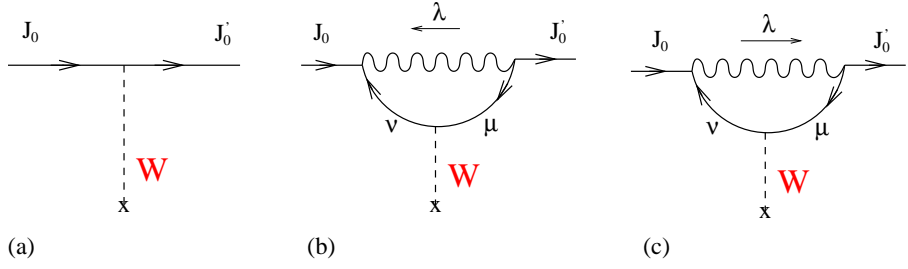}
\caption{Contributions to the weak matrix element of  $\langle
{\rm QP}| \hat W_{mean} |{\rm QP} \rangle$ type. \label{ris1}}
\end{center}
\end{figure}

$$
ME1=\langle {\rm QP} |  \hat W_{{\rm mean}} |{\rm QP} \rangle =
\langle \psi_0 | \alpha_{J_0} \alpha^\dagger_{\mu} \alpha_{\nu}
\alpha^\dagger_{J_0'} | \psi_0 \rangle \eta^{(+)}_{\mu \nu}
\overline{W}_{\mu \nu}=
$$
$$
=\Bigl( \delta_{J_0 \mu} \delta_{J_0' \nu} - \langle \psi_0 |
\alpha^\dagger_{\mu} \alpha^\dagger_{J_0'} \alpha_{J_0}
\alpha_{\nu}| \psi_0 \rangle \Bigr) \eta^{(+)}_{\mu \nu} \overline
{W}_{\mu \nu}.
$$
Using Eq. (\ref{3.3}) we can express the products of $\alpha_{J_0}
\alpha_\nu$ via phonon creation and annihilation operators
$Q^{\dagger}_\lambda$ and $Q_\lambda$:
$$
ME1=\eta^{(+)}_{J_0 J_0'} \overline{W}_{J_0 J_0'} + \sum_{\lambda}
(-1)^{J_0+J_\nu-J_\lambda} B^{\lambda}_{J_0 \nu} B^{\lambda}_{\mu
J_0'} \frac{2J_{\lambda}+1}{2J_0+1} \eta^{(+)}_{\mu \nu} \overline
{W}_{\mu \nu}.
$$
The first term corresponds to  Fig. \ref{ris1}a, while the second
term corresponds to  Fig. \ref{ris1}b. The sum over $\lambda$
includes all RPA solutions of Eq. (\ref{3.5}).   It is interesting
to recalculate this matrix element in a different way:
$$
\langle \psi_0 | \alpha_{J_0} \alpha^\dagger_{\mu} \alpha_{\nu}
\alpha^\dagger_{J_0'} | \psi_0 \rangle = \langle \psi_0 |
\alpha_{\nu} \alpha_{J_0} \alpha^\dagger_{\mu}
\alpha^\dagger_{J_0'} | \psi_0 \rangle .
$$
In this case the matrix element will be expressed as
$$
ME1=\sum_{\lambda} (-1)^{J_0+J_\nu-J_\lambda} A^{\lambda}_{J_0
\nu} A^{\lambda}_{\mu J_0'} \frac{2J_{\lambda}+1}{2J_0+1}
\eta^{(+)}_{\mu \nu} \overline{W}_{\mu \nu}
$$
that corresponds to  Fig. \ref{ris1}c.  Both calculations should
give the same result due to the completeness of RPA solutions. If
we leave only the most collective phonon in the sum, the result
will be different. In this case we prefer the first way of
calculation where the pure QP contribution appears explicitly. In
fact, this ambiguity is of minor importance since the weight of
the QP component in the wave function of the excited state is
rather small. The final result is:
$$
ME1(J_0, J_0')=\eta^{(+)}_{J_0 J_0'} \overline{W}_{J_0 J_0'} +
\sum_{\lambda} (-1)^{J_0+J_\nu-J_\lambda} B^{\lambda}_{J_0 \nu}
B^{\lambda}_{\mu J_0'} \frac{2J_{\lambda}+1}{2J_0+1}
\eta^{(+)}_{\mu \nu} \overline{W}_{\mu \nu}.
$$

\subsubsection{QP and QP + Phonon Mixing}

\begin{figure}
\begin{center}
\includegraphics[width=12cm]{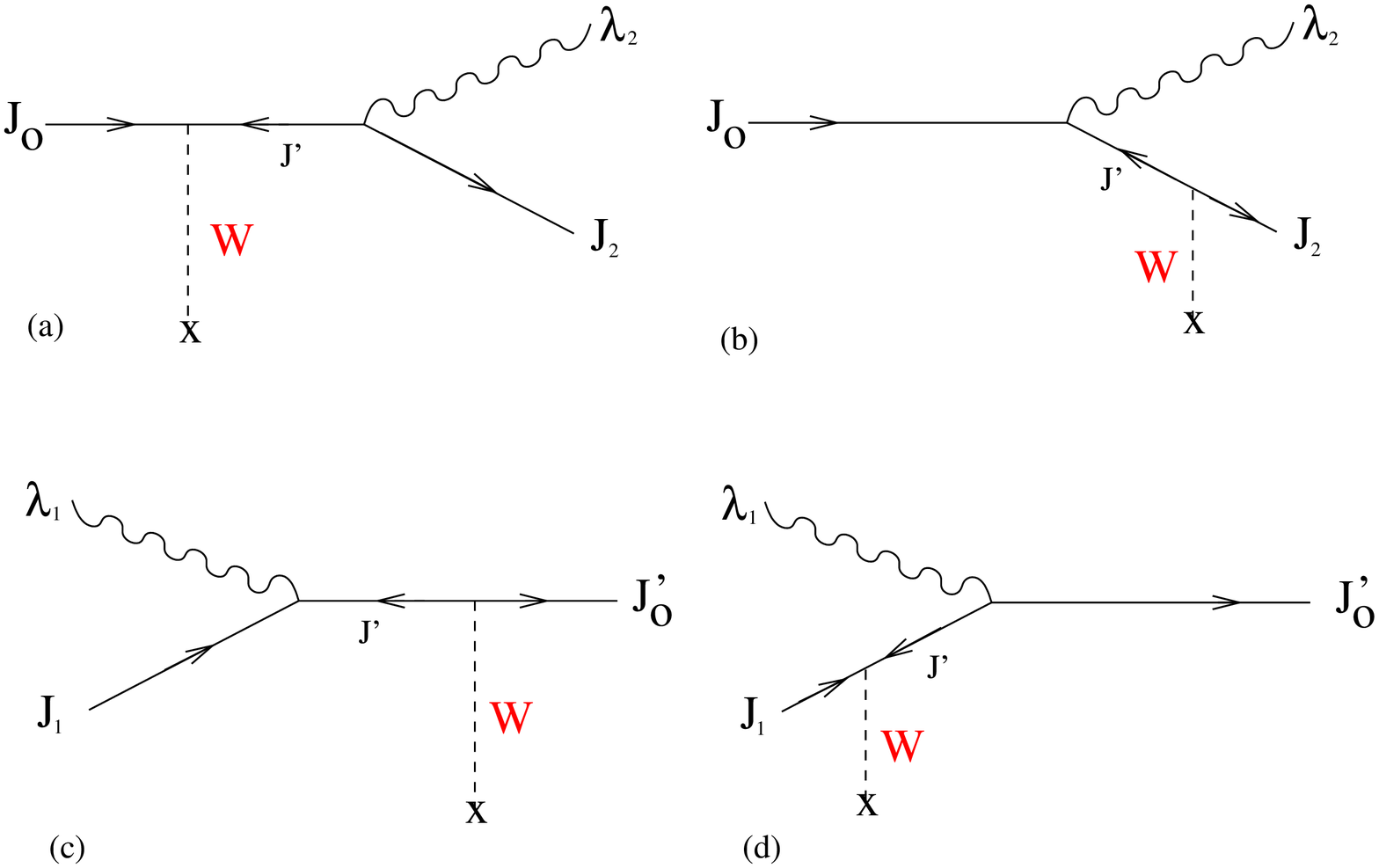}
\caption{Contributions to the weak matrix element of $\langle {\rm
QP} |\hat W_{{\rm mean}}|{\rm QP\,+\,phonon}\rangle$ and $\langle
{\rm QP\, +\, phonon}|\hat W_{{\rm mean}}|{\rm QP}\rangle$ types.
\label{ris2}}
\end{center}
\end{figure}
Now we calculate the weak matrix element between pure QP and
QP+phonon components:
$$
ME2 = \langle {\rm QP}|\hat W_{{\rm mean}}|{\rm QP\,+\,
phonon}\rangle =
$$
$$
=\langle \psi_0 | \alpha_{J_0} (\alpha_{\tilde \mu}
\alpha_{\nu}-\alpha^\dagger_{\mu} \alpha^\dagger_{\tilde \nu})
\alpha^\dagger_{J_2} Q^\dagger_{\lambda_2} | \psi_0 \rangle
\frac{1}{2} \xi^{(-)}_{\mu \nu} \overline{W}_{\mu \nu} C^{J_0
M_0}_{J_2 M_2 \; \lambda_2 \mu_2}.
$$
After simple transformations ME2 can be presented as follows:
$$
ME2(J_0, J_2 \lambda_2) =  \frac{\sqrt{2\lambda_2 +1}}{ \sqrt{2
J_0 +1}}  \sum_{J'} \left( \xi^{(-)}_{J_0 J'} \overline{W}_{J_0
J'} B^{\lambda_2}_{ J' J_2} - A^{\lambda_2}_{J_0 J'} \xi^{(-)}_{J'
J_2} \overline{W}_{J' J_2} \right).
$$
Here the first term corresponds to  Fig. \ref{ris2}a and the
second term corresponds to  Fig. \ref{ris2}b. Note the different
Bogolubov factors that appear here compared to those in the
previous matrix element.

\subsubsection{QP+Phonon and QP Mixing}
Here we calculate the similar weak matrix element between QP+phonon and
 QP components:
$$
ME3 = \langle {\rm QP\,+\,phonon}|\hat W_{{\rm mean}}|{\rm
QP}\rangle =
$$
$$
=\langle \psi_0 | Q_{\lambda_1} \alpha_{J_1} (\alpha_{\tilde \mu}
\alpha_{\nu}-\alpha^\dagger_{\mu} \alpha^\dagger_{\tilde \nu})
\alpha^\dagger_{J_0'} | \psi_0 \rangle \frac{1}{2} \xi^{(-)}_{\mu
\nu} \overline{W}_{\mu \nu} C^{J_0 M_0}_{J_1 M_1 \; \lambda_1
\mu_1}
$$
The resulting expression is:
$$
ME3(J_0', J_1 \lambda_1) =  (-1)^{\lambda_1 + J_1 + J_0}
\frac{\sqrt{2\lambda_1 +1}}{ \sqrt{2 J_0 +1}} \sum_{J'} \Bigl(
B^{\lambda_1}_{J_1 J'} \xi^{(-)}_{J' {J_0'}} \overline{W}_{J'
{J_0'}} -
 \xi^{(-)}_{J_1 J'} \overline{W}_{J_1 J'} A^{\lambda_1}_{J' J_0'} \Bigr).
$$
Again, the first term corresponds to  Fig. \ref{ris2}c and the
second term corresponds to Fig. \ref{ris2}d.

\subsubsection{QP+Phonon and QP+Phonon Mixing}

\begin{figure}
\begin{center}
\includegraphics[width=12cm]{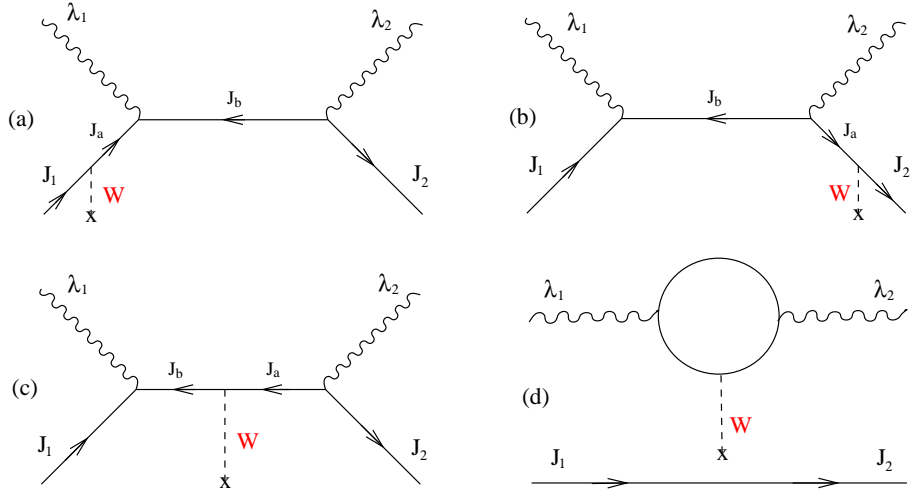}
\caption{Contributions to the weak matrix element of $\langle {\rm
QP\,+\,phonon}|\hat W_{{\rm mean}}|{\rm QP\,+\,phonon}\rangle$
type.                              \label{ris3}}
\end{center}
\end{figure}

There are several terms in this type of the weak matrix element.
Some of them are presented in Fig. \ref{ris3}; there are also
additional contributions with $\lambda_1 \leftrightarrow
\lambda_2$ and $\lambda_1 = \lambda_2$. Corresponding diagrams for
$\lambda_1 = \lambda_2$ are similar to those in Fig.
\ref{ris1}a,b.
$$
ME4 = \langle {\rm QP\,+\,phonon}|\hat W_{{\rm mean}}|{\rm
QP\,+\,phonon}\rangle=
$$
$$
\eta^{(+)}_{\mu \nu} \overline{W}_{\mu \nu} \langle \psi_0 |
Q_{\lambda_1} \alpha_{J_1} \alpha^\dagger_{\mu} \alpha_{\nu}
\alpha^\dagger_{J_2} Q^\dagger_{\lambda_2} | \psi_0 \rangle C^{J_0
M_0}_{J_1 M_1 \; \lambda_1 \mu_1}C^{J_0 M_0}_{J_2 M_2 \; \lambda_2
\mu_2}.
$$
It is convenient to split this matrix element into several terms,
$$
\langle \psi_0 | Q_{\lambda_1} \alpha_{J_1} \alpha^\dagger_{\mu}
\alpha_{\nu} \alpha^\dagger_{J_2} Q^\dagger_{\lambda_2} | \psi_0
\rangle= \delta_{J_1 \mu} \delta_{J_2 \nu} \langle \psi_0 |
Q_{\lambda_1}Q^+_{\lambda_2} | \psi_0 \rangle + \delta_{J_1 J_2}
\langle \psi_0 | Q_{\lambda_1} \alpha^\dagger_{\mu} \alpha_{\nu}
Q^\dagger_{\lambda_2} | \psi_0 \rangle-
$$
$$
- \delta_{J_1 \mu} \langle \psi_0 | Q_{\lambda_1}
\alpha^\dagger_{J_2} \alpha_{J_\nu}Q^\dagger_{\lambda_2} | \psi_0
\rangle - \delta_{J_2 \nu} \langle \psi_0 | Q_{\lambda_1}
\alpha^\dagger_{\mu} \alpha_{J_1}Q^\dagger_{\lambda_2} | \psi_0
\rangle + \langle \psi_0 | Q_{\lambda_1}
\alpha^\dagger_{\mu}\alpha^\dagger_{J_2}\alpha_{\nu}\alpha_{J_1}Q^\dagger_{\lambda_2}
| \psi_0 \rangle .
$$
The second term in this equation corresponds to Fig. \ref{ris3}d.
Here, the weak interaction  mixes $2^+$ and $2^-$ or $3^-$ and
$3^+$ phonon states. Since we do not include $2^-$ and  $3^+$
phonons in our consideration, we omit this term.

\underline{Part 1}:
$$
\delta_{J_1 \mu} \delta_{J_2 \nu} \langle \psi_0 |
Q_{\lambda_1}Q^\dagger_{\lambda_2} | \psi_0 \rangle \rightarrow
\delta_{\lambda_1 \lambda_2} \eta^{(+)}_{J_1 J_2}
\overline{W}_{J_1 J_2}.
$$
The result here is given by
$$
ME4.1(J_1 \lambda_1, J_2 \lambda_2)=\delta_{\lambda_1 \lambda_2}
\eta^{(+)}_{J_1 J_2} \overline{W}_{J_1 J_2}.
$$
The corresponding graph  is similar to  that of Fig. \ref{ris1}a,
where one should add a parallel phonon line and set  $J_1, J_2$
instead of $J_0, J_0'$ for the QP lines.

\underline{Part 2}:
$$
- \delta_{J_1 \mu} \langle \psi_0 | Q_{\lambda_1}
\alpha^\dagger_{J_2} \alpha_{J_\nu}Q^\dagger_{\lambda_2} | \psi_0
\rangle \approx - \delta_{J_1 \mu} \langle \psi_0 | Q_{\lambda_1}
[\alpha^\dagger_{J_2} \alpha_{J_\nu},Q^\dagger_{\lambda_2}] |
\psi_0 \rangle .
$$
After simple calculation one obtains
$$
ME4.2(J_1 \lambda_1, J_2 \lambda_2) =
\sqrt{(2\lambda_1+1)(2\lambda_2+1)} \sum_{J_a J_b} \eta^{(+)}_{J_1
J_a} \overline{W}_{J_1 J_a}
$$
$$
 \left( (-1)^{\lambda_2}(-1)^{J_1-J_b} A^{\lambda_2}_{J_a J_b} A^{\lambda_1}_{J_b J_2}
 \left\{
\begin{array}{rrr}
  \lambda_1 & J_b &  J_2    \\
  \lambda_2 & J_0 & J_1 \\
  \end{array}
\right\} +(-1)^{\lambda_1}(-1)^{J_1+J_b} B^{\lambda_1}_{J_a J_b}
B^{\lambda_2}_{J_b J_2} \frac{\delta_{J_b J_0}}{2J_0+1}
  \right).
$$
The second term here corresponds to  Fig. \ref{ris3}a, and the
first term corresponds to Fig. \ref{ris3}a with $\lambda_1
\leftrightarrow \lambda_2 $.

\underline{Part 3}:
$$
- \delta_{J_2 \nu}  \langle \psi_0 | Q_{\lambda_1}
\alpha^\dagger_{\mu} \alpha_{J_1}Q^\dagger_{\lambda_2} | \psi_0
\rangle \approx - \delta_{J_2 \nu} \langle \psi_0 | Q_{\lambda_1}
[\alpha^\dagger_{\mu} \alpha_{J_1},Q^\dagger_{\lambda_2}] | \psi_0
\rangle .
$$
The result is
$$
ME4.3(J_1 \lambda_1, J_2 \lambda_2) =
\sqrt{(2\lambda_1+1)(2\lambda_2+1)} \sum_{a b} \eta^{(+)}_{J_a
J_2} \overline{W}_{J_a J_2}
$$
$$
 \left( (-1)^{\lambda_2}(-1)^{J_1-J_b} A^{\lambda_2}_{J_1 J_b} A^{\lambda_1}_{J_b J_a}
 \left\{
\begin{array}{rrr}
  \lambda_1 & J_b &  J_2    \\
  \lambda_2 & J_0 & J_1 \\
  \end{array}
\right\} +(-1)^{\lambda_1}(-1)^{J_1+J_b} B^{\lambda_1}_{J_1 J_b}
B^{\lambda_2}_{J_b J_a} \frac{\delta_{J_b J_0}}{2J_0+1}
  \right).
$$
The second term here corresponds to  Fig. \ref{ris3}b, and the
first term corresponds to  Fig. \ref{ris3}b with $\lambda_1
\leftrightarrow \lambda_2 $. These terms are similar to those
calculated in Part 2.

\underline{Part 4}: This is the most complicated part of the
calculation.
$$
\langle \psi_0 | Q_{\lambda_1}
\alpha^\dagger_{\mu}\alpha^\dagger_{J_2}\alpha_{\nu}\alpha_{J_1}Q^\dagger_{\lambda_2}
| \psi_0 \rangle= \langle \psi_0 | \Bigl([Q_{\lambda_1},
\alpha^\dagger_{\mu}\alpha^\dagger_{J_2}]+\alpha^\dagger_{\mu}\alpha^\dagger_{J_2}Q_{\lambda_1}\Bigr)
\Bigl([\alpha_{\nu}\alpha_{J_1},Q^\dagger_{\lambda_2}] +
Q^\dagger_{\lambda_2}\alpha_{\nu}\alpha_{J_1}\Bigr)|\psi_0
\rangle=
$$
$$
=\langle \psi_0 | \Bigl(A^{\lambda_1}_{\mu J_2} C^{ \lambda_1
\mu_1}_{J_\mu M_\mu \; J_2 M_2}
+\alpha^\dagger_{\mu}\alpha^\dagger_{J_2}Q_{\lambda_1}\Bigr)\Bigl(
A^{\lambda_2}_{J_1 \nu } C^{ \lambda_2 \mu_2}_{J_1 M_1 \; J_\nu
M_\nu} +
Q^\dagger_{\lambda_2}\alpha_{\nu}\alpha_{J_1}\Bigr)|\psi_0
\rangle=
$$
$$
=A^{\lambda_1}_{\mu J_2} A^{\lambda_2}_{J_1 \nu } C^{ \lambda_1
\mu_1}_{J_\mu M_\mu \; J_2 M_2}C^{ \lambda_2 \mu_2}_{J_1 M_1 \;
J_\nu M_\nu}+ \zeta C^{ \lambda_3 -\mu_3}_{J_\mu M_\mu \; J_2 M_2}
C^{ \lambda_4 -\mu_4}_{J_1 M_1 \; J_\nu M_\nu} B^{\lambda_3}_{\mu
J_2} B^{\lambda_4}_{J_1 \nu } \langle \psi_0 |Q_{\lambda_3}
Q_{\lambda_1}Q^\dagger_{\lambda_2}Q^\dagger_{\lambda_4}|\psi_0
\rangle.
$$
Here
$$
\zeta =(-1)^{\lambda_3 +\mu_3+\lambda_4 +\mu_4},
$$
$$
\langle \psi_0 |Q_{\lambda_3}
Q_{\lambda_1}Q^\dagger_{\lambda_2}Q^\dagger_{\lambda_4}|\psi_0
\rangle= \delta_{\lambda_1 \lambda_2}\delta_{\lambda_3 \lambda_4}+
\delta_{\lambda_1 \lambda_4}\delta_{\lambda_2 \lambda_3},
$$
and we use the RPA approximations
$$
[Q_{\lambda_1},
\alpha^\dagger_{\mu}\alpha^\dagger_{J_2}]=A^{\lambda_1}_{\mu J_2}
C^{ \lambda_1 \mu_1}_{J_\mu M_\mu \; J_2 M_2} + O(\alpha^\dagger
\alpha) \approx A^{\lambda_1}_{\mu J_2} C^{ \lambda_1
\mu_1}_{J_\mu M_\mu \; J_2 M_2},
$$
$$
[\alpha_{\nu}\alpha_{J_1},Q^\dagger_{\lambda_2}]=A^{\lambda_2}_{J_1
\nu } C^{ \lambda_2 \mu_2}_{J_1 M_1 \; J_\nu M_\nu} +
O(\alpha^\dagger \alpha) \approx A^{\lambda_2}_{J_1 \nu } C^{
\lambda_2 \mu_2}_{J_1 M_1 \; J_\nu M_\nu}.
$$
After some calculations we obtain
$$
ME4.4(J_1 \lambda_1, J_2 \lambda_2) =
\sqrt{(2\lambda_1+1)(2\lambda_2+1)} \sum_{a b} \eta^{(+)}_{J_a
J_b} \overline{W}_{J_a J_b}
$$
$$
\left(
 (-1)^{ \lambda_2}(-1)^{J_1 - J_b}
 A^{\lambda_2}_{J_1 J_b} A^{\lambda_1}_{J_a J_2}
\left\{
\begin{array}{rrr}
  \lambda_1 & J_b &  J_2    \\
  \lambda_2 & J_0 & J_1 \\
  \end{array}
\right\}+ (-1)^{\lambda_1}(-1)^{J_1+J_b} B^{\lambda_1}_{J_1 J_b }
B^{\lambda_2}_{J_a J_2} \frac{\delta_{J_b J_0}}{2J_0+1}\right)+
$$
$$
+\delta_{ \lambda_1 \lambda_2} \sum_{\lambda \nu \mu}
(-1)^{J_1+J_\nu-J_\lambda} B^{\lambda}_{J_1 \nu} B^{\lambda}_{\mu
J_2} \frac{2J_{\lambda}+1}{2J_1+1} \eta^{(+)}_{\mu \nu}
\overline{W}_{\mu \nu}.
$$
The second term corresponds to  Fig. \ref{ris3}c, and the first
term corresponds to  Fig. \ref{ris3}c with $\lambda_1
\leftrightarrow \lambda_2 $. The last term corresponds to Fig.
\ref{ris1}b with the additional parallel phonon line and $J_1,
J_2$ instead of $J_0, J_0'$ for the QP lines.

\end{document}